\documentclass[11pt]{article}
\usepackage{mathptmx}

\begin{document}

\title{Quantum corrections to finite-gap solutions for  Yang-Mills-Nahm equations via zeta-function technique.}
\author{Sergey Leble\\Gdansk University of Technology, \\{\small
ul. Narutowicza 11/12, 80-952, Gdansk, Poland,
}}
\maketitle
\date{\today}

\begin{abstract}
One-dimensional Yang-Mills-Nahm  models are considered from  algebrogeometric points of view.
 A quasiclassical quantization of the models based
on path integral and its zeta function representation in terms of a
Green function diagonal for a heat equation with an elliptic
potential  is considered. The Green function diagonal and, hence, zeta function and its derivative are expressed via solutions of Hermit equation and, alternatively, by means of  Its-Matveev formalism in terms of  Riemann teta-functions.
     For the  Nahm model, which field is  represented via elliptic
(lemniscate) integral by construction, one-loop quantum corrections to action are evaluated  as the zeta function derivative in zero point in terms of a hyperelliptic
integral. The alternative expression should help to link the representations and continue investigation of the Yang-Mills-Nahm  models.

\end{abstract}

\section{  Introduction. On Nahm models}

The celebrated Yang-Mills field theory  have
reductions to one dimensional models \cite{Nahm}.
There are two possibilities of such reduction interpretation:

\textbf{Euclidean version;} the equations reduces  by deleting dependence on arbitrary
three cartesian coordinates,
  intimately related   the
 Atiyah-Drinfeld-Hitchin-Manin- Nahm (ADHMN)  construction to static
monopole solutions to
 Yang-Mills-Higgs  theories in four
dimensions in the
 Bogomolnyi-Prasad-Sommerfield
limit. The ADHMN construction supply equivalence
between  self-dual equations, one - unidimensional, the other in
three dimensions (reduced  Euclidean four dimensional theory by
deleting dependence on a single variable) \cite{Corr}.

 \textbf{Minkowski space}:  ought to  delete
dependence on spatial variables: \cite{Bas}.

Let us consider \textbf{Yang-Mills equations} in four Euclidean dimensions
 \begin{equation}\label{YM}
   D_{\mu}T_{\mu\nu}=0,
\end{equation}
for the gauge fields $T_{\mu} = T_{\mu}^+,\; \mu=1,2,3,4 $, where
$$
T_{\mu\nu} =T_{\nu,\mu} - T_{\mu,\nu}-\imath[T_{\mu},T_{\nu}],
\quad D_{\mu}\Phi = \partial_{\mu} - \imath [T_{\mu},\Phi].
$$
Suppose independence on  variables $x_k$, k=1,2,3;  setting $x_4 =x$,
\begin{equation}\label{YM1}
\begin{array}{c}
     \frac{d^2T_k}{dx^2} = [T_j[T_j,T_k]], \quad
  [T_k,\frac{d T_k}{dx}]=0. \\
\end{array}
\end{equation}
The  self-dual  equations,
\begin{equation}\label{SDE}
    \frac{dT_i}{dx}=\pm \varepsilon_{ijk}T_jT_k,
\end{equation}
 imply Eqs. (\ref{YM1}).

This paper   develops results of recent publication of the author \cite{LE3},
which mathematical origin strictly relates to the pioneering Its-Matveev paper \cite{IM}.

In the Sec. 2 we review and specify main definitions of the theory,  describing briefly zeta function formalism of quasiclassical approximation for path integral representation of the quantum field amplitudes. The quantum corrections to static solutions of a reduced Nahm model, expressed in
  terms of elliptic integral are evaluated via Riemann zeta function of heat kernel operator Green function diagonal. The Sec. 3 starts with explicit description of a static solutions of the reduced Nahm model, it continues the  formalism description introducing an equation for the Laplace transform of the Green function diagonal.  The equation is related to the Hermit equation \cite{C.H}, it is solved algebraically. The inverse Laplace transform is given by hyperelliptic (g=2) integral. The Mellin transformation finalize the zeta function derivation, its derivative give the formula for quantum correction. The fourth section is devoted to Matveev-Its formula application to the spectral problem which appear within the variables division procedure applied to the mentioned Green function problem. Its diagonal values, integrated by spectrum, give alternative expression for the Riemann zeta function and correspondent one-loop corrections. The resulting expression is formed by well-converged theta-functions series which allows its effective numerical evaluation.

\section{On quasiclassical quantum corrections in path integral formalism.}
\subsection{ Yang-Mills Lagrangian for the Nahm reduction.}

To show more details of the description we restrict ourselves by the case 2x2 matrices $T_i$ (isospin 1/2 case)  and scalar field (zero spin) with respect to usual spin classification.

 \textbf{Reductions of Nahm system}.
There are two possibilities

1. Let $\sigma_i$ be Pauli matrices, simple  substitution  $T_i=\phi_i(x)\sigma_i$ gives the Euler system for
$\phi_i(x)$   solved in Jacobi functions \cite{LE1}.

2. The ansatz  in (\ref{SDE}) with upper sign
\begin{equation}\label{ansatz}
   T_i =\phi(x)\alpha_i , i=l,2,3,
\end{equation}
where  $\alpha_i$ - constant matrices, yields \cite{Corr}
    \begin{equation}\label{nonSDE}
   \begin{array}{c}
       2 \alpha_i = \sum_{j=1}^{3}[\alpha_j[\alpha_j,\alpha_i]]      \\
   \phi''(x)=2\phi^3, \\
   \end{array}
 \end{equation}
 to be considered in this paper.
  The equation for $\phi(x)$ enters
a specific  class  (m=0) of
 Ginzburg-Landau model  with the potential function of the field
\begin{equation}\label{V}
	V(\phi)=\phi^4/2+c.
\end{equation}
Let us choose the variable $x$ as the proper time and take the Weyl gauge $T_4=0$ as in \cite{Corr}, $T_i=\phi(x)\sigma_i$ from (\ref{ansatz}), then
\begin{equation}\label{T0i}
T_{0i}= T_{i,0}= \phi'(x)\sigma_i;
\end{equation}
more general case of the $T_i$ choice is investigated in \cite{Corr}. Next, for $i,k=1,2,3,$
\begin{equation}\label{Tik}
T_{ik}= T_{i,k}-T_{k,i}-\imath[T_i,T_k]= -\imath \phi^2(x)[\sigma_i,\sigma_k]=2\phi^2(x)\epsilon_{ikl}\sigma_l.
\end{equation}
Hence the Lagrangian density is proportional to  (see, e.g. \cite{FP})
\begin{equation}\label{Lagr}
T_{\mu\nu}T^{\mu\nu}=T_{0i}T^{0i}+T_{ik}T^{ik}=  (\phi'(x))^2\sigma_i\sigma_i+4\phi^4(x)\epsilon_{ikl}\sigma_l\epsilon_{ikl'}\sigma_{l'}=3[(\phi'(x))^2+
8\phi^4(x)].
\end{equation}
A quasiclassical quantization is based on the path integral formalism  for YM theory \cite{FP}.

\subsection{Quantum corrections via zeta-function}

The  fields $\phi(x)$ of the classical theory may be considered as stationary or static solutions of    the nonlinear Klein-Gordon-Fock equation
$$
\phi_{tt}-\phi_{xx} +
V^{\prime\prime}(\phi)=0,
$$
with the operator
\begin{equation}\label{V''}
\label{D}D = - \partial^{2}_{x}   +
V^{\prime\prime}(\phi(x)).
\end{equation}
The approximate quantum corrections  to the solutions
equation  are obtained by path integral estimated by stationary phase method analog. For a one-dimensional potential $V(\phi(x))$,
the operator $D=-\frac{d^2}{dx^2} + u(x)$. appears
  while the second variational derivative
of the  action functional (Lagrange density $\mathcal{L}%
=(\mathbf{\partial}\phi)^{2}/2  - V(\phi)$) is evaluated.

 The quantum correction takes the form (for details and refs. see \cite{LE3})
 \begin{equation}
\label{qucor1}\Delta S_{qu} =S_{qu} - S_{vac}= \frac{ \hbar}{2}\ln(\frac{\det{D}}{\det{D_{0}}%
}).
\end{equation}
Some comments on vacuum contribution $S_{vac}$ one can find in Appendix.

A link to the diagonal  Green function
(\emph{heat kernel formalism}) has been used in quantum theory
since works by Fock \cite{F}.
We  study the problem  for a periodic u(x), the Green function is defined via:
$$\left(\frac{\partial}{\partial t} + D\right) g_{D}\left(t,x,x_{0}\right) =\delta\left(t \right)\delta\left(x-x_{0}\right),$$
where  $D=-\frac{d^2}{dx^2} + u(x)$, $u(x) \leq u_m$ and $
 \lambda \in [\lambda_0, +\infty) $. Eventual account of other variables (see again \cite{Bas}) is shown in Appendix.
Let the set $\{\lambda_{n}\} = S$ be a spectrum of a linear
operator $D$, then
  the  generalized
Riemann zeta-function $\zeta_{D}(s)$ is defined by the equality
\begin{equation}
\label{zet0}\zeta_{D}(s) = \sum_{\lambda_{n}\in
S}\lambda_{n}^{-s},
\end{equation}
as
analytic continuation to
the complex plane of $s$ from the half plane  in
which the sum in (\ref{zet0}) converges. Differentiating the
relation (\ref{zet0}) with respect to $s$ at the point $s=0$
yields
\begin{equation}
\label{lndet} \ln(\det D)=
\zeta_{D}^{\prime}(0).
\end{equation}
The  zeta-function (\ref{zet0}) admits the
representation via the diagonal of the Green function  $g_{D}(t,x,x_0)$ of the
operator $\partial_{t}+D$.

Define the function
\begin{equation} \label{defg}
 \begin{array}{c}
   \gamma_{D}\left(t\right)=\int g_{D}\left(t,x,x\right)dx, \\
  \end{array}
 \end{equation}
which  Mellin transformation gives  zeta function of the operator D
  \begin{equation}\label{defZ}
    \zeta_{D} (s)=\frac{1}{\Gamma(s)} \int_{0}^{+\infty}t^{s-1}\gamma_{D}(t)dt.
\end{equation}

\section{Hermite equation for a Green function diagonal and its solutions}

\subsection{Static solutions of the reduced Nahm model}

Integral of (\ref{nonSDE})
   \begin{equation}\label{Wphi4}
(\phi')^2 = (\phi^2)^2 -b^4
\end{equation}
may be considered as the case m=0 of the Ginzburg-Landau model.  The fact that the YM Lagrangian is reduced to the form (\ref{Lagr}) by simple rescaling of $\phi$ allows to link the mass of a Nahm field particle with the quantum correction to action evaluated below. Due to (\ref{Wphi4}),   the solution of Nahm equation is expressed as inversion of the elliptic (lemniscate)
integral
\begin{equation}\label{dphi}
   \int_0^\phi \frac{d\phi}{\sqrt{\phi^4-b^4}} = \frac{1}{b}\int_{0}%
^{\frac{\phi}{b}\ }\frac{dt}{\sqrt{\left(  t^{2}-1\right)  \left(
t^{2}+1\right)  }}=x,\,
\end{equation}
which gives the Jacobi $sn$ function with the imaginary modulus $k=i$
\begin{equation}\label{wsn}
 \phi=bsn(ibx,i).%
\end{equation}

\subsection{Generalized Hermite equation}

 One can check by direct substitution, that the Laplace transform $\hat{g}_L(p,x,x_0)$ of the Green function   diagonal
   $G(p,x) = \hat{g}_L(p,x,x)$ is a solution of bilinear  equation
\begin{equation}\label{Hermit}
    2GG'' - (G')^2 - 4(u(x)-p)G^2+1=0,\qquad
\end{equation}
which is of the same origin as Hermit equation \cite{C.H} (see also \cite{BBEIM}).
In a case of reflectionless and finite-gap solutions  the equation (\ref{Hermit}) is
solved more effectively than the equation for the Green function.
Namely, the  polynomials  (in p) $P,Q$
\begin{equation}\label{PQ}
    G(p,x)=P(p,)/2\sqrt{Q(p,z)},
\end{equation}
where the new variable
\begin{equation}\label{z}
	z=cn^2(bx;i)
\end{equation}
 is introduced (for a compactness of expressions $b\rightarrow -ib$ is redefined).

It yields solutions of (\ref{Hermit}) by
 plugging (\ref{PQ}):
\begin{equation}\label{linkPQ}
    b^2(\rho(2PP''-(P')^2)+\rho'PP')-(p+u(z))P^2+Q=0,
\end{equation}
the primes denote derivatives with respect to z, while  account of the definition of $u(x)=6\phi^2$ in (\ref{V''}),  explicit expression for $V$
in (\ref{V}), the form of solution (\ref{wsn}) and the new variable expression (\ref{z}) should have in mind.
 \begin{equation}\label{Nahm}
 \begin{array}{c}
                   \rho =         z(1-z)(2-z),   \\
                u(z) =   -6b^{2}\left( 1-z\right).                \\
                            \end{array}
 \end{equation}

In our case of linear  $u(z)$, which corresponds the Nahm model (\ref{Nahm}), the choice is given by
\begin{equation}\label{PQD}
\begin{array}{c}
 P=p^2+P_1(z)p+P_2(z),  \\
  Q=p^5+q_4p^4+q_3p^3+q_2p^2+q_1p+q_0. \\
\end{array}
\end{equation}
 Plugging (\ref{PQD}) into (\ref{linkPQ}) yields (argument z is not shown)
\begin{equation}\label{splitD}
    \begin{array}{c}
       -2P_1  - u  + q_4 = 0, \\
    -2P_2  - P_1^2  - 2 u P_1 +b^2(2\rho P_1''+\rho'P_1')+q_3=0,    \\
       b^2(\rho (2P_2''+2P_1P_1''-(P_1')^2)+\rho'(P_2'+P_1P_1'))-2P_1P_2)-u(2P_2+P_1^2)+q_2 =0,\\
       b^2(2\rho(2P_1''P_2-P_1'P_2'+P_1P_2'')+\rho'(P_1P_2'+P_1'P_2))-P_2^2 - 2uP_1P_2 + q_1=0,\\
       b^2(\rho(2P_2P_2''-P_2'^2)+\rho'P_2P_2')-uP_2^2+q_0=0.\\
     \end{array}
   \end{equation}

 The substitution of (\ref{Nahm}) into (\ref{splitD}) gives the values of the polynomial Q coefficients
  $ q_{4}=0,\quad q_{3}=
-21b^{4}, \quad $ $ q_{2}=
q_{1}=\,108b^{8}, q_{0}= 0, $ hence $P_1(z)=
-3b^2(z-1), \quad P_2=18b^4z^2-36b^4z.$
  Finally,
 \begin{equation}\label{Q}
    Q=\prod_{i=1}^{i=5}(p-p_i),
\end{equation}
where the polynomial $Q$ have the simple  roots $p_i=\{-2\sqrt{3}b^2,- 3b^2,0,3b^2,2\sqrt{3}b^2\}$,    ordered  for real $b^2$; obvious reflection symmetry implies an underlying Riemann surface reduction \cite{BBEIM}.

Let us pick up the expressions determining $\hat{\gamma}(p)$:
\begin{equation}\label{hatgammap}
    \hat{\gamma}(p) = \int( p^2
-3b^2(z-1)p+18b^4(z^2-2z))dx/2\sqrt{Q}.
 \end{equation}
Going to the variable z  and integrating from z=1 to z=0 gives the Laplace transform
 \begin{equation}\label{hatgammap'}
\begin{array}{c}
 \hat{\gamma}(p)=  [K(\imath)p^{2}-3b^{2}%
(K(\imath)-E(\imath))p-12b^{4}K\left(  \imath\right)  ] /2\sqrt{Q},\\
\end{array}
   \end{equation}
 As a next step we obtain
 \begin{equation}\label{gamat}
\gamma(t)=\int	\exp[-pt]\hat{\gamma}(p)dp,
\end{equation}
 as inverse Laplace transform.
The Mellin transformation
 gives the zeta function   in terms of complete elliptic
lemniscate integrals $K(\imath)$ and $E(\imath)$, that are expressed, for example,
as hypergeometric series \cite{Erd}.

Plugging the result (\ref{hatgammap'}) into the Riemann zeta function (\ref{defZ}), denoting for brevity $K(\imath)=K, E(\imath)=E$, yields
\begin{equation} \label{zetaa}
\begin{array}{c}
\zeta(s)=
 - \int_{l}\frac{1}{(-p)^{s}}\frac{2Kp^{2}+3b^{2}%
(K-E)p-48b^{4}K  }{2\sqrt{p(p+3b^{2})(p-3b^{2}%
)(p-2\sqrt{3}b^{2})(2\sqrt{3}b^{2}+p)}}dp.
\end{array}%
\end{equation}
The change of the variable of integration $p=3p'b^2, dp = 3dp'b^2$ gives
\begin{equation}\label{zet}
\zeta(s)=\ \frac{b}{2}\int_{l} \frac{  4K \ -3p^{2}K  +3p[K  -E ]  }{(-3b^{2}p)^{s}\sqrt{p\left(  3p^{4}-7p^{2}+4\right)  }}ds.
\end{equation}

Differentiation by s and going to the limit $s\rightarrow 0$,
 gives the integral that is proportional to
 the mass correction (Nahm particle mass):
 \begin{equation}\label{zeprimze}
\zeta^{\prime}(0)=\ \frac{b}{2}\int_{l\ }\frac{  4K\ -3p^{2}%
K+3p[K-E]  }{\sqrt{p\left(  3p^{4}-7p^{2}+4\right)  }}\ln\left(
-3b^{2}p\right)  ds.
\end{equation}

The integral (\ref{zeprimze}) corresponds the  hyperelliptic curve of genus 2
\begin{equation}\label{curve}
	\mu^2=p(p-1)(p+1)(p-2/\sqrt{3})(p+2/\sqrt{3})
\end{equation}
 that is a particular case $\alpha=-1,\beta=2/\sqrt{3} $ of an example from Sec. 7.1 of the book \cite{BBEIM}. The important property of the curve, the presence of non-trivial authomorphism T is described:
  \begin{equation}
	T: (\mu,p)\stackrel{}{\rightarrow}(\frac{\mu(-\beta^{3/2})}{p^3},\frac{-\beta}{p}).
\end{equation}
As it is shown in  the mentioned book the properties of the curve (\ref{curve}) allow to transform the integral by hyperelliptic curve to the combination of elliptic ones by the correspondent curves.

\section{ Alternative construction. The generalized zeta-function via Its-Matveev formula.}

The one-dimensional Green
function in the heat kernel formalism is defined by
\begin{equation}\label{g1}
    \left(\frac{\partial}{\partial t} - \frac{d^2}{dx^2}+ V'' \right) G\left(t,x,x_{0}\right) =\delta\left(t \right)\delta\left(x-x_{0}\right)
\end{equation}
    with a  potential $V''(\phi(x))=6b^2sn^2(\imath bx,\imath)$ originated from a model like (\ref{nonSDE}), see also (\ref{wsn}).
 The function $G(t,x,x_0)$ is supposed to be continuous at $t \geq 0$.
The division of variables
\begin{equation}\label{division}
G(t,x,x_0)=\int g_H(x,x_0)\exp[Ht]dH,	
\end{equation}
yields the spectral problem
\begin{equation}\label{a}
    \left( H - \frac{\partial^2}{\partial x^2} +6b^2sn^2(\imath bx,\imath)\right)\psi\left(x,H \right)=0,
\end{equation}
or, by rescaling $y=\imath bx,  h=\frac{H}{b^2} $, results in the Jacobi form of Lam$\acute{e}$ equation \cite{Erd} (n=2, k=i)
 \begin{equation}\label{aa}
    \left(h + \frac{\partial^2}{\partial y^2}-6(\imath sn(y,i))^2\right)\psi\left(y,h\right)=0.
\end{equation}
The   Its-Matveev formula \cite{IM} reads as
\begin{equation}\label{IMpsi}
\psi\big(y;h(\mathcal{P})\big)=\mathrm{const}(\mathcal{P})
\frac{\Theta\big( \mathcal{A} (\mathcal{P})+y U +D\big)}
{\Theta\big(yU+D\big)}e^{\Omega(\mathcal{P})y}.
\end{equation}
It is expressed in terms of Riemann theta functions and solves the
  equation (\ref{aa}) for  a spectral
parameter h with the potential
\begin{equation}\label{IMpot}
u=-2 \frac{ d^2}{ dy^2}\ln\Theta\big( U y + D\big)+const.
\end{equation}
  The parameters $\mathcal{P},U,D,\mathcal{A},\Omega(\mathcal{P}) $ are defined via Abelian integrals fixed by positions of singular points in (\ref{Q}), see Sec. 7.7 \cite{BBEIM}, where the reduction to Lam$\acute{e}$ potential case (\ref{aa}) is specified.
 The  theta function of the representation   is defined, e.g. in the same book \cite{BBEIM} via theta-series.
 The series convergence is rapid, therefore the representation
  (\ref{IMpsi}) is convenient for numeric
evaluation of the integrals in zeta formalism.
The link of the potentials   for the case  of the Sine-Gordon model (Lam$\acute{e}$ equation with n=1)
is expressed in explicit form in \cite{LE3}.

 The Green function $g_h$ of the spectral Lam$\acute{e}$ problem (\ref{aa}) may be constructed as a product of two
independent solutions $\psi_{+}(y;h)$, and $\psi_-$ of the spectral equation with the same h:
\begin{equation}\label{g_h}
    g_h(y,y_0)=\frac{1}{ W} \{\begin{array}{c}
                                            \psi_{+}(y;h)
  \psi_{-}(y_0;h),\quad y<y_0\\
        \psi_{-}(y;h)
  \psi_{+}(y_0;h)                                   \quad y>y_0.\end{array}
  \end{equation}
The Wronskian factor $W$ is chosen to normalize
(\ref{Gryd}) so that fix the jump of the first derivative
with respect to x:
\begin{equation}\label{g_h}
    \lim_{\epsilon \rightarrow 0}[\frac{dg_H(x_0+\epsilon,x_0)}{dx}
    -
\frac{dg_h(H_0-\epsilon,x_0)}{dx}]=-1.
\end{equation}
The independent solution $\psi_{-}(y;h) $ may be chosen
antisymmetric with respect to the reflection $x \rightarrow -x$.

The result (\ref{g_h}) is substituted into the integral by the
spectrum
\begin{equation}\label{Gryd}
  G(t,x,x_0) =  b^2 \int
   g_h(\imath bx,\imath bx_0)\exp[ h b^2 ]dh.
\end{equation}

 Finally we integrate the diagonal values of the Green function from (\ref{Gryd})
by the period of a solution obtaining  $\gamma_{D}(t)=\int G(t,x,x)dx$.
After that, using the definition  of the generalized zeta-function   (\ref{zeta}) and
calculating  the derivative at zero point,  one
arrives at the mass expression that is proportional to the quantum correction (\ref{qucor1}).

\section{Conclusion}
Two alternative expressions (\ref{zet}) and one arising from  (\ref{Gryd}) for the zeta function link the Its-Matveev representation and the Laplace transform (hyperelliptic integral) representation. We study both representations of the zeta function because of eventual significance of the result, its cross-verification and, the necessity of further investigation of both ones. Perhaps, there is some interest of such comparison from mathematical point of view,   The value of zeta function derivative at zero point  (\ref{zeprimze}) allows to evaluate the quantum corrections to energy that may be considered as the YMN particle mass itself (zero value of classical mass) in the framework of the Nahm model as YM reduction, after the choice of the parameter b, that also need additional  physical investigation.

Dressing procedure \cite{Mat,Mat2} may be applied directly to the spectral problem (\ref{a}) or to the evolution equation (\ref{g1})  and  widen class of solutions. Some such results concern 2+1 case \cite{Nahm, LE2}.
The YM equations itself is proved to be non-integrable, but the dressing (gauge-Darboux transformation of \cite{LE2}) and methods outlined in this article
are effective in the case of YM reductions as well as for other non-integrable systems as e.g. Landau-Ginzburg one.

\section{Appendix: model implementation in multidimensions and regularization}
Let us consider a problem in d dimensions for a Nahm field that still
depends only on x but linkwed to the ADHMN construction.
\begin{equation}
\label{D}D = - \partial^{2}_{x} - \Delta_{y} +
V^{\prime\prime}(\phi(x))=D_x- \Delta_{y}.
\end{equation}
  $y \in R^{d-1}$ stands for a transverse variables.
For a regularization
  a vacuum
action $S_{vac}$ is introduced with
$ D_{0} = -
\partial^{2}_{x} - \Delta_{y}$. The quantum correction takes the form
 \begin{equation}
\label{qucor1}\Delta S_{qu} =S_{qu} - S_{vac}= \frac{ \hbar}{2}\ln(\frac{\det{D}}{\det{D_{0}}%
}).
\end{equation}
Such extraction is used when the limit case of a kink is studied \cite{LE3}.

For the transversal variables contribution one can use
   the  property of
multiplicity: \textit{if the operator D is a sum of two
differential operators $D = D_{1} + D_{2}$, which depend on
different variables, the following equality holds}
\begin{equation}
\label{gg}\gamma_{D}(t)=\gamma_{D_{1}}(t)\gamma_{D_{2}}(t).
\end{equation}
It follows directly from definition  (\ref{defg}) of  $\gamma_{D}(t)$.
For the Laplacian $\Delta_{y}$ the function $\gamma$ is equal to
$d-1$-dimensional Poisson integral
\begin{equation}
\label{Poi}
\gamma_{D_y}(t) = \frac{1}{(2\pi)^{d-1}}\int_{R^{d-1}%
}d\mathbf{k} \exp(-|\mathbf{k}|^{2}t) = (4\pi t)^{- \frac{d-1}{2}}.
\end{equation}
Similarly the "vacuum" one is evaluated.	
Substitution the expression (\ref{Poi}) having in mind (\ref{gamat}) for $\gamma(t)$
 \begin{equation}\label{zeta}
\zeta_D(s)=\frac{1}{\Gamma(s)}
\int_{0}^{+\infty}t^{s-1}\left(\gamma_{D_{0}}(t)-(4\pi t)^{- \frac{d-1}{2}}\gamma(t)\right)dt
\end{equation}
yields the regularized zeta function
 \begin{equation}\label{zeta}
\zeta (s)=\frac{1}{\Gamma(s)}
\int_{0}^{+\infty}t^{s-1}\left(\gamma_{D_{0}}(t)-\gamma_{D}(t)\right)dt.
\end{equation}

Finally, the quantum correction to action is proportional to
\begin{equation}
 \Delta S=\frac{\hbar}{2}\zeta' (0).
\end{equation}

\end{document}